\documentclass[twocolumn,showpacs,preprintnumbers,amsmath,amssymb]{revtex4}

\usepackage{graphicx}
\usepackage{dcolumn}
\usepackage{bm}

\begin{document}

\title{ Calculated effects of disorder on the Mo core levels in purple bronze Li$_2$Mo$_{12}$O$_{34}$.}

\author{T. Jarlborg}

\affiliation{
DQMP, University of Geneva, 24 Quai Ernest-Ansermet, CH-1211 Geneva 4,
Switzerland (retired)
}

\date{\today}

\begin{abstract}
The band structures of ordered and thermally disordered Li$_2$Mo$_{12}$O$_{34}$ are calculated by use
of ab-initio DFT-LMTO method with focus on the behavior of the Mo 3d-core levels. 
It is shown that thermal disorder and zero-point motion lead to substantial core level broadening, and the broadening
at room temperature is predicted to be sufficiently larger than at zero degrees to allow for a detection
by XPS measurements. However, real purple bronze has 10 percent of Li vacancies and static disorder will attenuate the $T$-dependent
broadening. It is argued that core level spectroscopies could be a useful tool for measuring of thermal disorders in many materials,
especially for those with minor static disorder. 
Studies of core levels in magnetic materials will be helpful for an understanding of $T$-dependent spin moments.

\end{abstract}

\pacs{71.23.-k,75.10.-b,78.30.Ly}

\maketitle

Keywords: Electronic structure, thermal disorder, core-level spectroscopy.

\section{Introduction.}

The Lithium Purple Bronze, (Li$_{0.9}$Mo$_{6}$O$_{17}$) is
an unusual material which draws attention to research for nearly
three decades \cite{McC,Schlen}. 
It is a compound with a
rather complicated layered structure \cite{onoda} and it exhibits
highly 1-dimensional (1D) electronic
properties \cite{green,whang,popo}. For example, characteristic 1D features have been seen
in scanning tunnel spectroscopy \cite{hage,podl}, from magnetoresistance \cite{xu}, but also dimensional crossover from 
thermal expansion \cite{sant} and even the role of disorder has been invoked for a low-$T$ transition \cite{chak}.
Superconductivity below 2K has also been reported \cite{sant,chak}.
 Indeed band calculations
starting with 3D electronic interactions reveal an electronic
structure with 1D character \cite{whang,popo}, with large band
dispersion only in one of the directions within the layers.
The band structure shows that two
bands are very close together near the crossing of the Fermi energy, $E_F$,
with a moderately large density-of-states (DOS).
Thermal disorder and possible spin fluctuations on the 
valence bands lead to band broadening and partial gaps at $E_F$ \cite{jcg}.


\begin{figure}
\includegraphics[height=6.0cm,width=8.0cm]{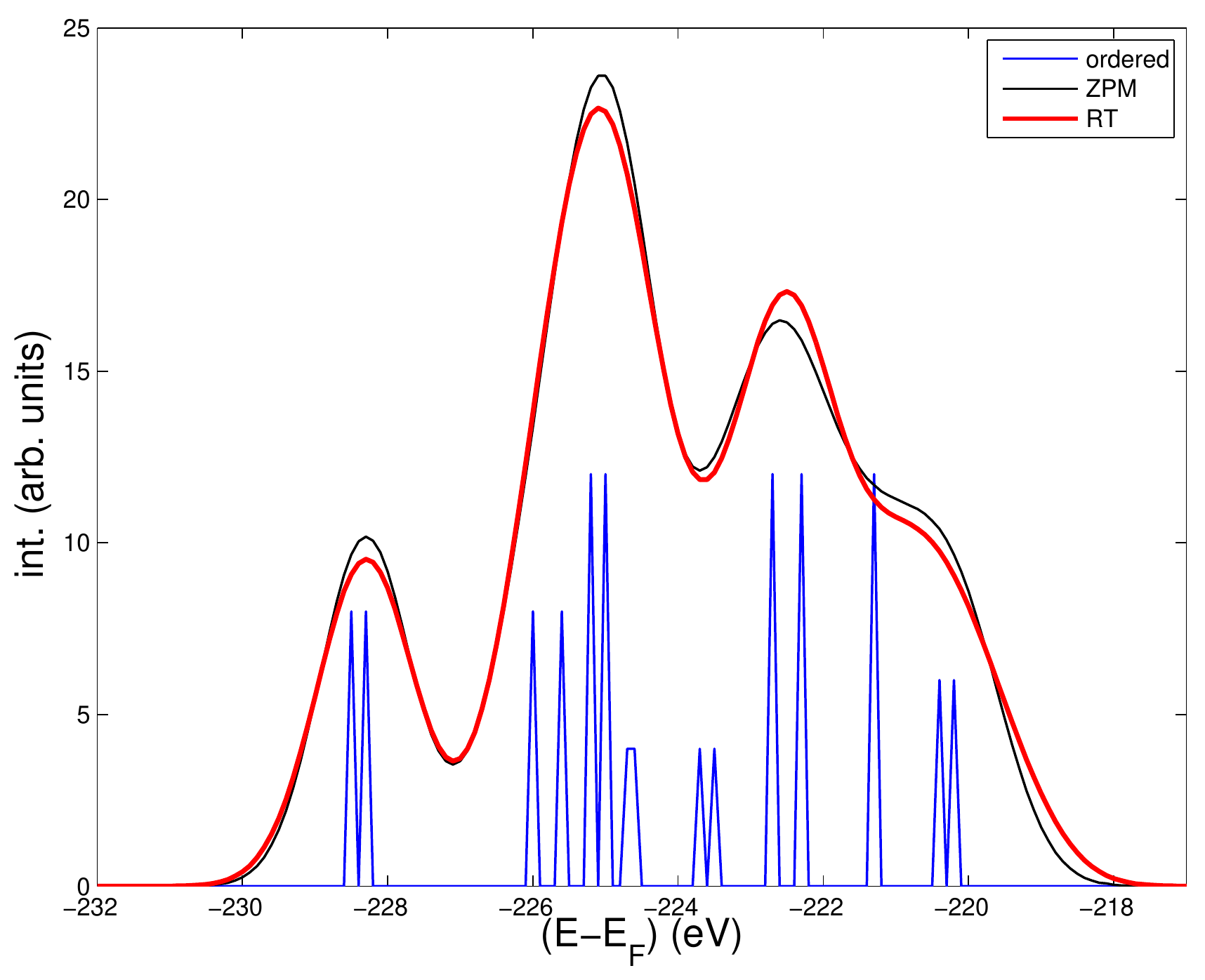}
\caption{Calculated core Mo-3d levels in the undistorted cell of Li$_2$Mo$_{12}$O$_{34}$
(blue thin lines). Intensities from all Mo-3d levels for 4 disordered structures at low $T$ "ZPM"
(thin black line),
and for 5 disordered structures corresponding to RT (bold red line).
A Gaussian broadening of 1.5 eV is added in the ZPM and RT cases to
account for static disorder. The effect from the same broadening on the core levels in the ordered cell
is shown in Fig. \ref{figure2}} \label{figure1}
\end{figure}


Several band calculations have been made for the ordered lattice of stoichiometric
purple bronze
\cite{whang,popo,meri,nuss}. Band structure calculations rarely take into
account thermal distortions of the lattice positions. However,
structural disorder due to thermal vibrations are detectable through spectroscopic methods \cite{jcg,roshed,wilk,dug,opt}, and it is recognized as
being important for $T$-dependent physical properties in materials with particular fine structures
in the DOS near $E_F$
\cite{jcg,fesi,ped,ce,dela,giust,marini,h3s,h3sb}. Here for purple bronze, effects of thermal
fluctuations at large $T$ and zero-point motion (ZPM) at $T=0$ might be pertinent for the degree of dimensionality and
the band overlap between the two bands at $E_F$ \cite{jcg}. The present work show that core level broadening from thermal disorder
can be large, and therefore it is important to recognize core level spectroscopy as a useful tool for determination of effects
of disorder on valence states and material properties.

\section{Calculations.} \label{sec:band}

\subsection{Thermal disorder and zero-point motion.} \label{sec:disorder}

The band structure is sensitive to disorder (and ZPM) because of the fluctuations of the
potential in a vibrating disordered lattice.
The Coulomb potential $v_i(r)$ at a point $r$ within a site $i$ is
\begin{equation}
v_i(r) = - \sum_j Z_j/|r-R_j| + \int_0^{\infty} \rho(r')/|r-r'| d^3r'
\end{equation}
where $Z_j$ are the nuclear charges on sites $j$, $\rho(r)$ is the electron charge density,
and the sum and integral cover all space.
The contribution to $v_i(r)$ from its own site (with radius $S_i > r'$) can be separated from the
contribution from the surrounding lattice through the technique of Ewald lattice summation \cite{zim}:
\begin{equation}
v_i(r) \approx -Z_i/r + \int_0^{S_i} \rho(r')/|r-r'|d^3r' + M_i
\label{eqV}
\end{equation}
Thus, the Coulomb interaction with the outside lattice is condensed into a Madelung shift, $M_i$ \cite{arb}. 
\begin{equation}
M_i = \sum_j m_{i,j} (Z_j -q_j + 2\pi S_j^2 \rho)/a_0
\label{mij}
\end{equation}
where $m_{i,j}$ is the geometric structure factor, 
$q_j$ is the electron charge on site $j$ and $\rho$ is the interstitial charge density.

This shift is
 identical for equivalent sites if the lattice is perfectly ordered. But all sites have different $M_i$
in a disordered lattice, partly because of the local differences in atomic positions and partly
because of the charge transfers induced by the disorder. Thus,
the potentials at different sites are slightly different and they
vary in time. Phonons are very slow compared to the electronic time scale and the electronic
structure can relax adiabatically.

Phonons are excited thermally following the
Bose-Einstein occupation of the phonon DOS,
$F(\omega)$.
The averaged atomic displacement amplitude, $\sigma$, can be calculated
as function of $T$ \cite{zim,grim}.  The result is approximately
that $\sigma_Z^2 \rightarrow 3\hbar\omega_D/2K$ at low $T$ due to ZPM
and $\sigma_T^2 \rightarrow 3 k_BT/K$ at high $T$
(``thermal excitations''), where $\omega_D$ is a weighted average of $F(\omega)$. The force constant, $K=M_A\omega^2$,
where $M_A$ is an atomic mass (here the mass of Mo is used because of its
dominant role in the electronic DOS), can be calculated as
$K = d^2E/du^2$ ($E$ is the total energy), or it can be taken from experiment.
We use the measurements
of the phonon DOS of the related blue bronze K$_{0.3}$MoO$_3$ \cite{requ} to estimate $K$ and the
average displacements of Mo atoms, as is explained in ref. \cite{jcg}.

The individual displacements $u$ follow a Gaussian distribution function;
\begin{equation}
g(u) = (\frac{1}{2\pi\sigma^2})^{3/2} exp(-u^2/2\sigma^2)
\label{Neffeq}
\end{equation}
where $\sigma$, the standard deviation of $u$, will be a parameter in the
different sets of calculations. In order to estimate the
effect of such atomic displacements on the band structure, each
atomic site in the unit cell is assigned a random displacement ($u$)
along $x,y$ and $z$ following the Gaussian distribution function.
Our calculations are made for supercells that contain such ($T$-dependent) distortions.
 The disorder make the potentials different on different sites
in the fully self-consistent calculations, and the core levels on different sites
are no longer degenerate as in an ordered supercell. The average spread of the core level values
defines the energy broadenings, $\Delta\epsilon$, see later.

The parameter $\sigma$ depends on $T$ and the
properties of the material. From the experimental data in blue
bronze \cite{requ} we estimate the $T$-dependence of $\sigma$ for
purple bronze, so that $\sigma_{Z}/b_0$ ($b_0$ is the $y$-axis lattice constant \cite{xyz}) is of the
order 0.7 percent for Mo, and thermal vibrations become larger
than $\sigma_Z$ from about 120 K.
Band calculations are made for a total of nine different
disordered configurations. Four of these configurations have a $\sigma/b_0 < 0.7$ percent,
and the remaining have larger $\sigma$, to represent disorder just below room temperature (RT).

 The
electronic structure of Li$_2$Mo$_{12}$O$_{34}$ (two formula units
of stoichiometric purple bronze) has been calculated using the
Linear Muffin-Tin Method (LMTO, \cite{oka,arb,bdj}) in the local
density approximation \cite{lda} (LDA), with or without
disorder \cite{jcg}. The lattice dimension and atomic
positions of the structure have been taken from Onoda {\it et al}
\cite{onoda}. The lattice constant, $b_0$, in the conducting
$y$-direction is 5.52 \AA ~and 
along the least conducting $x$-direction it is $12.76$ \AA. In order to adapt the
LMTO basis for an open structure as purple bronze we inserted 56
empty spheres in the most open parts of the structure. This makes
totally 104 sites within the unit cell. The basis consists of
s-,p- and d-waves for Mo, and s- and p-states for Li, O and empty
spheres, with one $\ell$ higher for the 3-center terms.
Corrections for the overlapping atomic spheres are included. All
atomic
sites are assumed to be fully occupied (except for the case with a
vacant Li atom, see later), and they
are all considered as inequivalent in the calculations.
Self-consistency is made using 125 k-points.
Other details can be found 
in ref. \cite{jcg}. Valence energies converge more rapidly than core levels
and the total energy. It was necessary to continue the iterations
for some of the disordered configurations in \cite{jcg} in order to have stable
core levels, even though the the valence state were converged already.

\subsection{Static disorder}\label{sec:stat-disor}

In addition to the thermal effects, on which we concentrated in
the previous section, disorder can come from a variety
of other sources. In particular, imperfections of the crystal (from
non-stoichiometry, vacancies, site exchange, and twins in single crystals) should 
contribute to a $T$-independent disorder. In particular, one can worry about the fact that
one out of ten Li atoms are missing in the real material. The unit
cell considered here contains two Li, and the influence on the
electronic bands from the replacement of one of these with an
empty sphere (without taking structural distortion into account)
is moderately large \cite{jcg}. The additional 1.5 eV Gaussian broadening
in Fig. \ref{figure1} is an effort to account for such static disorders, but it is a free parameter
that has to be refined when results are to be compared with experiment. 

 In order to get an idea of the effects of Li vacancies we also calculate the
electronic structure of LiMo$_{12}$O$_{34}$ (one missing Li in 2 formula units
of the normal undistorted cell). All other details of the calculation are unchanged.

\subsection{Calculated results}

 The band dispersion
along the conducting $\Gamma-Y$ (or $P-K$) direction agrees
well with the measured results obtained by ARPES \cite{wang} showing a
flattening of the two dispersive bands at about 0.4-0.5 eV below
$E_F$. One important result of ref. \cite{jcg} was the suggestion that thermal
disorder makes a broadening of the valence band, with an amplitude comparable
to the separation of the two bands near $E_F$. Thus, the broadening explains
 why only one band is observed in APRES.  The overall agreement
between different band results and photoemission is reassuring for
this complicated structure. Here, we will focus on the
Mo 3d core level and their dependence on disorder.

\begin{table}[ht]
\caption{\label{table1}
Energies $(\epsilon)$ relative to $E_F$ of the upper SO-split Mo 3d-core levels 
and the average energy shifts $\Delta\epsilon_{ZPM}$ and $\Delta\epsilon_{RT}$ from ZPM and thermal disorder at RT,
respectively. The lower of the SO-split levels are ~3.3 eV more bound. All energies in eV.}
  \vskip 2mm
  \begin{tabular}{l c c c}
  \hline
     level  & $\epsilon$  & $\Delta\epsilon_{ZPM}$  & $\Delta\epsilon_{RT}$ \\

  \hline \hline

Mo far from Li (oct. O)   & -222.36 &  0.14 & 0.23 \\
Mo far from Li (oct. O)   & -222.73 &  0.10 & 0.12 \\
Mo intermed. (oct O)   & -225.02 & 0.13  & 0.27 \\
Mo intermed. (oct O)   & -225.22 & 0.17  & 0.34 \\
Mo near Li (tet O)     & -221.36 & 0.23 & 0.44 \\
Mo near Li (tet O)     & -220.33 & 0.25 & 0.66 \\

  \hline
  \end{tabular}
\end{table}

The Mo sites can be divided into 3 groups according to their local $N(E_F)$ and
proximity to Li. The first group contains the four most distant Mo from Li,
see ref. \cite{onoda}. They have a large local DOS at $E_F$
and determine the two bands that cross $E_F$ \cite{jcg}. The second group has intermediate distances
to the Li-atoms, and their DOS are quite low. Each Mo in these two groups are surrounded by
octahedrons of oxygens, as can be seen in ref. \cite{onoda}. The last group contains the 4 Mo closest
to Li, with practically no local DOS at $E_F$, and each Mo is surrounded by a tetrahedron of oxygens.
Within each group there are slight differences in the core levels, see Table \ref{table1}, because the local
differences in the structure make the Madelung potential different
among the groups of Mo sites. 
The spread in core levels $\Delta\epsilon_{ZPM}$
and $\Delta\epsilon_{RT}$ in Table \ref{table1} is based on the variations of core levels
on the different sites among the 4 disordered "ZPM" and 5 "RT" configurations, respectively.
The statistics is limited, but it appears clear that the Mo near Li are most
sensitive to disorder. This makes sense, because the screening is weakend by the low local DOS.
Oppositely, on the high-DOS Mo's in the first group, the spread of core levels is lowest.
The core levels in the last group, those closest to Li, are least bound, and they make up the
rightmost peak of the calculated intensities in Figure \ref{figure1} from the ordered structure,
the ZPM- and the RT-configurations, respectively. Thus the 3 groups of Mo sites have quite separate
3d-core levels, and the peak to the right due to Mo's near Li is most affected by disorder.
Measurements at low and high $T$ should therefore be most likely seen on the right hand side
of the spectrum.

\begin{figure}
\includegraphics[height=6.0cm,width=8.0cm]{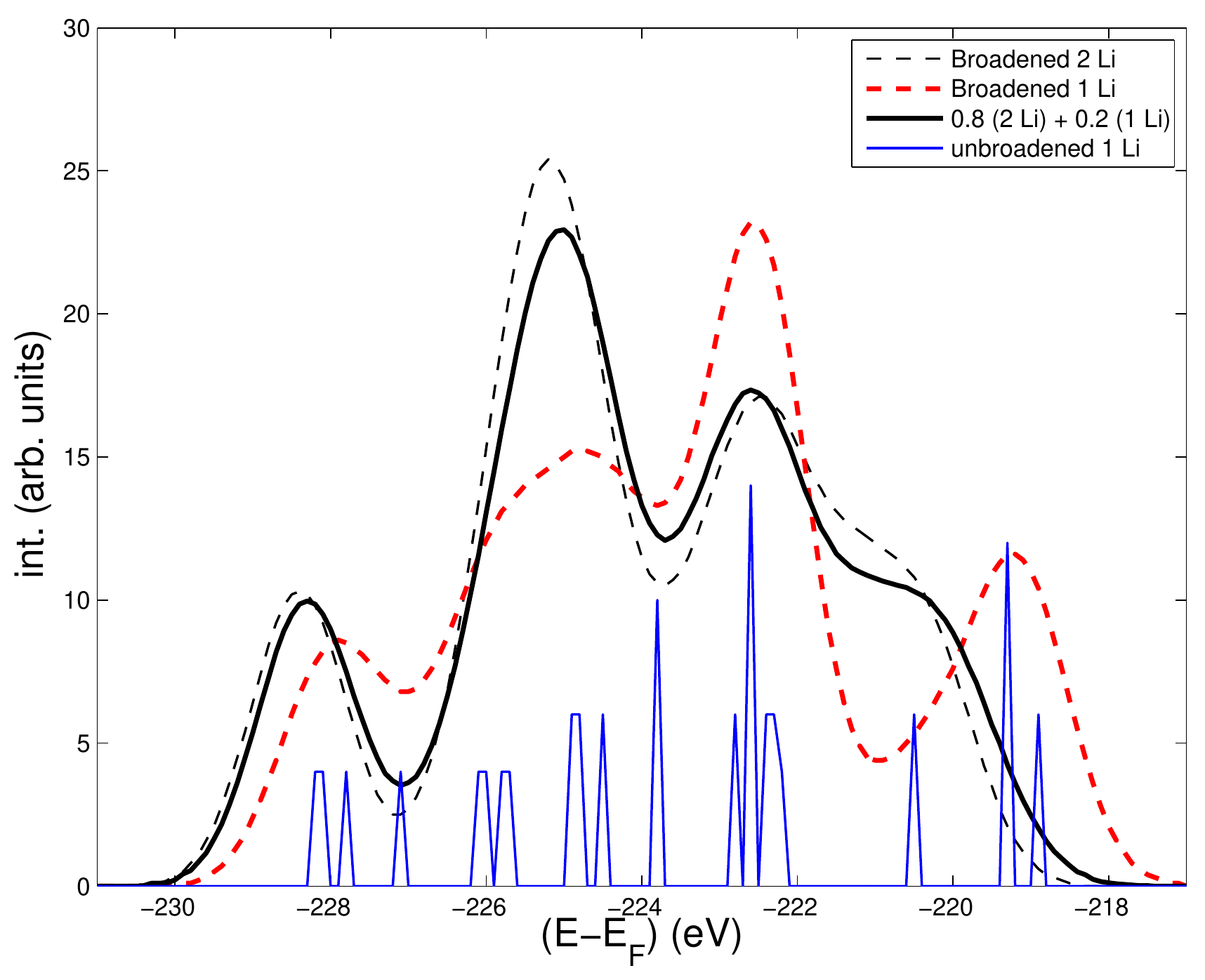}
\caption{Energy distribution of the Mo-3d core levels (including a Gaussian broadening
of about 1.5 eV)  
in  Li$_2$Mo$_{12}$O$_{34}$ (black broken thin line), 
for LiMo$_{12}$O$_{34}$ (red broken bold line) and a superposition of the two 
previous curves with weights 0.8 and 0.2, respectively. The core levels in 
LiMo$_{12}$O$_{34}$ without Gaussion broadeneing is shown by the thin (blue) lines.} \label{figure2}
\end{figure}

The intensities in Fig \ref{figure1} contain all the spin-orbit
split 3d levels, and a Gaussian broadening of 1.5 eV
to account for static disorder. Purple bronze has 10 percent Li vacancies,
which most likely induce structural deformations around the vacancies. Other lattice
defects and non-stoichiometric conditions will also deform the structure locally.
All together this adds up to a static disorder, which will modify the core levels
by an unknown amount. 

 In order to get an idea about the sensitivity of the electronic structure to Li-vacancies we made
 a calculation where one of the Li is missing, but where the structure
is undistorted compared to the ideal structure. The resulting core level intensities
are shown in Figure \ref{figure2}. As seen, the general shape
of the Mo 3d levels is different from the one for stoichiometric case in
Figure \ref{figure1}. The least bound 4s levels on Mo near Li are changing most, which is normal
since one Li is missing. Those Mo levels tend to be more
isolated (increasing their energies) from the other levels. There are also some changes for
the relative intensities at intermediate
binding (Mo far from Li). A simple weighting of the two distributions (with adjustment
of the Fermi levels to account for one less electron in the Li deficient case)
is shown by the bold (red) line.
The core levels could also be modified by lattice
relaxation around the missing Li. The sensitivity of the core level energies
to a single Li vacancy indicates that some differences between the calculated
and the measured shapes of the intensities can come from static disorders, but the strength of the T-independent
broadening (in addition to that of ZPM) is not known. 
A Gaussian broadening of more than 1 eV is probably needed to account for the static disorder.
However, an analyze of the relative T-dependent variations of calculated and measured intensities
can still be made independently of the initial shape of the spectrum.



\section{Conclusion.}

The calculations show $T$-dependent broadening of the 3d core levels on Mo. However, static disorder due to Li vacancies
interfere with the $T$-dependent effects. The degree
of disorders differ at inequivalent Mo-sites and lead to  
larger broadening for sites near the Li. Note that the assumed structural data for the stoichiometric perfect
unit cell in the calculations are coming from measurements on real purple bronze with defects,
and that the measured Mo anisotropy parameter
is largest for the Mo near Li \cite{onoda}. 
Another source for discrepancy between experimental and theoretical T-dependences could come from different
distortion amplitudes $u$, since $\sigma$ is extrapolated from measurements in blue bronze \cite{requ}. 
 These results on core level broadening give support to the conclusion \cite{jcg} that it is difficult to observe the band
separation in high precision ARPES \cite{wang,wang1,denl}, and the broadening from static and vibrational disorder
might also mask the signature of Luttinger behavior near $E_F$ \cite{wang,wang1,gweon}. 

With all these precautions for purple bronze it is suggested that $T$-dependent studies of core levels would be more revealing
for other materials where static imperfections are unlikely. It can also be interesting to study magnetic T-dependent
or pressure ($P$) dependent transitions through studies of core levels. In theory, it has been shown that the spin-splitting
of core levels in fcc Cerium will change drastically at the $(T,P)$-dependent $\alpha - \gamma$ transition \cite{ce,jmg}.
The magnetic phase has a wide spread of individual spin polarized core levels, because of local thermally induced lattice distortions.

Acknowledgement. I am grateful to J.D. Denlinger for useful discussions about preliminary experimental results.
This work was presented at the ACSIN2016 conference in Frascati, Italy, October 9-13, 2016, 
"Atomically Controlled Surfaces, Interfaces and Nanostructures" edited by Antonio Bianconi and Augusto Marcelli 
(Superstripes Press, Rome, Italy) 2016  ISBN 9788866830597.


\begin{thebibliography}{10}


\bibitem{McC} W.H. McCarroll and M. Greenblatt, J. Solid State Chem. {\bf 54}, 282, (1984).

\bibitem{Schlen} C. Schlenker, H. Schwenk, C. Escribe-Filippini, and J. Marcus, Physica {\bf B}135, 511, (1985).

\bibitem{onoda} M. Onoda, K. Toriumi, Y. Matsuda and M. Sato, J. Solid State Chem. {\bf 66}, 163, (1987).

\bibitem{green} M. Greenblatt, W.H. McCarroll, R. Neifeld, M. Croft and J.V. Waszczak,
Solid State Commun. {\bf 51}, 671, (1984).

\bibitem{whang} M.-H. Whangbo and E. Canadell, J. Am. Chem. Soc. {\bf 110}, 358, (1988).

\bibitem{popo} Z.S. Popovi\'{c} and S. Satpathy, Phys. Rev. B{\bf 74}, 045117, (2006).

\bibitem{hage} J. Hager, R. Matzdorf, J. He, R. Jin, D. Mandrus, M.A. Cazalilla and E.W. Plummer, Phys. Rev. Lett. {\bf 95}, 186402, (2005).

\bibitem{xu} X. Xu, A.F. Bangura, J.G. Analytis, J.D. Fletcher, M.M.J. French, N. Shannon, J. He, S. Zhang, D. Mandrus, R. Jin
   and N.E. Hussey,  Phys. Rev. Lett. {\bf 102}, 206602, (2009).
   
\bibitem{podl} T. Podlich, M. Klinke, B. Nansseu, M. Waelsch, R. Bienert, J. He, R. Jin, D. Mandrus and R. Matzdorf, J. Phys.: Condens. 
Matter {\bf 25}, 014008, (2013).   

\bibitem{sant} C.A.M. dos Santos, B.D. White, Y-K. Yu, J.J. Neumeier and J.A. Souza, Phys. Rev. Lett. {\bf 98}, 266405, (2007).

\bibitem{chak} J. Chakhalian, Z. Salman, J. Brewer, A. Froese, J. He, D. Mandrus and R. Jin, Physica {\bf B} 359-361, 1333, (2005).

\bibitem{jcg} T. Jarlborg, P. Chudzinski and T. Giamarchi, Phys. Rev. B{\bf 85}, 235108, (2012).

\bibitem{meri} J. Merino and R.H. McKenzie, Phys. Rev. B{\bf 85}, 235128, (2012).

\bibitem{nuss} M. Nuss and M. Aichhorn, Phys. Rev. B{\bf 89}, 045125, (2014).

\bibitem{roshed} L. Hedin and A. Rosengren, J. Phys. F: Metal Phys. {\bf 7}, 1339, (1977).

\bibitem{wilk} R.H. McKenzie and J.W. Wilkins, Phys. Rev. Lett. {\bf 69}, 1085, (1992).

\bibitem{dug} S.B. Dugdale and T. Jarlborg, Solid State Commun. {\bf 105}, 283, (1998).

\bibitem{opt} T. Jarlborg, Phys. Rev. B{\bf 76}, 205105, (2007).

\bibitem{fesi} T. Jarlborg, Phys. Rev. B{\bf 59}, 15002, (1999).

\bibitem{ped} P. Pedrazzini, H. Wilhelm, D. Jaccard, T. Jarlborg, M. Schmidt, M. Helfland, L. Akselrud,
H.Q. Yuan, U. Schwarz, Yu. Grin and F. Steglich, Phys. Rev. Lett. {\bf 98}, 047204, (2007).

\bibitem{dela} O. Delaire, K. Marty, M.B. Stone, P.R.C. Kent, M.S. Lucas, D.L. Abernathy,
D. Mandrus and B.C. Sales, PNAS {\bf 108}, 4725, (2011).

\bibitem{ce} T. Jarlborg, Phys. Rev. B{\bf 89}, 184426, (2014) DOI:10.1103/PhysRevB.89.184426.

\bibitem{giust} F. Giustino, S.G. Louie and M.L. Cohen, Phys. Rev. Lett. {\bf 105}, 265501, (2010).

\bibitem{h3s} T. Jarlborg and A. Bianconi, Scientific Reports {\bf 6}, 24816, (2016).

\bibitem{h3sb} A. Bianconi and T. Jarlborg, Europhysics Letters {\bf 112}, 37001, (2015). 

\bibitem{marini} E. Cannuccia and A. Marini, Phys. Rev. Lett. {\bf 107}, 255501, (2011).

\bibitem{zim} J.M. Ziman, {\it Principles of the Theory of Solids} (Cambridge University
Press, New York, 1971).

\bibitem{arb} T. Jarlborg and G. Arbman , J. Phys. F{\bf 7}, 1635, (1977).

\bibitem{grim} G. Grimvall, {\it Thermophysical properties of materials.}
(North-Holland, Amsterdam, 1986).

\bibitem{requ} H. Requardt, R. Currat, P. Monceau, J.E. Lorenzo, A.J. Dianoux, J.C. Lasjaunias and
J. Marcus, J. Phys.: Condens. Matter {\bf 9}, 8639, (1997).

\bibitem{xyz} We adopt the same conventions as in refs. \cite{onoda,popo} for the structure, site-notation
and k-space, so that $\vec{x}$ and $\vec{z}$ are parallel and $\vec{y}$ perpendicular to the planes.

\bibitem{oka} O.K. Andersen, Phys. Rev. B{\bf 12}, 3060, (1975).

\bibitem{bdj} B. Barbiellini, S.B. Dugdale  and T. Jarlborg, Comput. Mater. Sci. {\bf 28}, 287, (2003).

\bibitem{lda} W. Kohn and L.J. Sham, Phys. Rev. {\bf 140}, A1133, (1965).

\bibitem{wang} F. Wang, J.V. Alvarez, J.W. Allen, S.-K. Mo, J. He, R. Jin, D. Mandrus and H. H\"{o}chst,
Phys. Rev. Lett. {\bf 103}, 136401, (2009).

\bibitem{wang1} F. Wang, J.V. Alvarez, S.-K. Mo, J.W. Allen, G.H. Gweon, J. He, R. Jin,
D. Mandrus and H. H\"{o}chst,
Phys. Rev. Lett. {\bf 96}, 196403, (2006).

\bibitem{denl} L. Dudy, J.D. Denlinger, J.W. Allen, F. Wang, J. He, D. Hitchcock, A. Sekiyama, 
and S. Suga, J. Phys. Condensed Matter {\bf 25}, 014007 (2013).

\bibitem{gweon} G.H. Gweon, J.D. Denlinger, J.W. Allen, C.G. Olson, H. Hoechst, J. Marcus and C. Schlenker, Phys. Rev. Lett.
{\bf 85}, 3985, (2000).

\bibitem{jmg} T. Jarlborg, E.G. Moroni and G. Grimvall, Phys. Rev. B{\bf 55}, 1288, (1997).

\end{thebibliography}
\end{document}